# One-dimensional Topological Edge States of Bismuth Bilayers

Ilya K. Drozdov[1*], A. Alexandradinata[1*], Sangjun Jeon[1], Stevan Nadj-Perge[1], Huiwen Ji[2], R. J. Cava[2], B. A. Bernevig[1], and Ali Yazdani[1]

[1]Joseph Henry Laboratories & Department of Physics, Princeton University, Princeton, NJ 08544, USA

[2]Department of Chemistry, Princeton University, Princeton, NJ 08544, USA

* These authors contributed equally to this work

The hallmark of a time-reversal symmetry protected topologically insulating state of matter in two-dimensions (2D) is the existence of chiral edge modes propagating along the perimeter of the system[1–5]. To date, evidence for such electronic modes has come from experiments on semiconducting heterostructures in the topological phase which showed approximately quantized values of the overall conductance[6–8] as well as edge-dominated current flow[9]. However, there have not been any spectroscopic measurements to demonstrate the one-dimensional (1D) nature of the edge modes. Among the first systems predicted to be a 2D topological insulator are bilayers of bismuth (Bi)[4] and there have been recent experimental indications of possible topological boundary states at their edges[10–13]. However, the experiments on such bilayers suffered from irregular structure of their edges or the coupling of the edge states to substrate's bulk states. Here we report scanning tunneling microscopy (STM) experiments which show that a subset of the predicted Bi-bilayers' edge states are decoupled from states of Bi substrate and provide direct spectroscopic evidence of their 1D nature. Moreover, by visualizing the quantum interference of edge mode quasi-particles in confined geometries, we demonstrate their remarkable coherent propagation along the edge with scattering properties that **are consistent with strong suppression of backscattering as predicted for the propagating topological edge states.**

Early theoretical consideration of electronic model of graphene with spin-orbit coupling[1] provided the impetus for the realization of topological states of matter which can occur in two[14] and three-dimensions[15,16]. A freestanding bilayer of Bi (in the (111) plane of the Bi rhombohedral structure[17]) can be regarded as buckled honeycomb structure (Figure 1a), which, unlike graphene, possesses strong spin-orbit coupling and represents an elemental 2D topological insulator model system[4,18]. Theoretical modeling of the electronic structure of Bi-bilayer on various substrates shows that while electronic hybridization modifies the properties of this system, a subset of topological edge modes should be present for supported bilayers[12,19]. Remarkably, a bulk crystal of Bi along the [111] direction can be considered as a stack of Bi-bilayers that are weakly bonded to each other[17] (Fig. 1a). When considering the edges of an atomically ordered Bi-bilayer on the surface of a Bi-crystal, it is clear that the edges can have different geometries and therefore couple differently to the substrate. In general, it is possible to have armchair or zigzag edges with the zigzag edges (as shown in Fig. 1a) terminated such that atoms are either close to the vacuum (type A) or to the substrate underneath (type B). If such edges were stabilized and examined experimentally, we could determine whether they possess topological edge modes and also understand how the coupling to the bulk states might modify them. Our model calculations (see supplementary section I for our calculation and related work by others) show that while hybridization at type B edges with bulk states of Bi substrate strongly suppresses edge state signatures, a subset of topological edge mode at type A edge does not hybridize as much and should be detectable in STM studies.

Fortuitously, cleaving single crystals of Bi in the (111) plane we find surfaces that are terminated by Bi-bilayer islands with remarkably straight and disorder-free edges. STM images of such surfaces show that all the islands have the expected 4Å height for Bi-bilayers (Fig. 1b) and all the extended edges are identified to be of the zigzag type (see atomically resolved topographic image of the edge on the inset of Fig. 1b). STM spectroscopic measurements of Bi-bilayer islands show that their electronic structure away from their edges can be understood based on the Bi(111) surface state band structure. Spectra in the middle of islands (gray line Fig. 1c) show a symmetric peak centered at +213 meV that can be associated with the saddle point in the Bi(111) surface state band structure[17,20,21] (with logarithmic energy dependence around +213 meV, see supplementary section II).

At their edges, Bi-bilayer islands show spectroscopic features that demonstrate the propagation of 1D electrons along the edges, as well as the sensitivity of such edge states to coupling to the bulk. Spectroscopic measurements of a large number of islands reveal two distinct categories of spectra at their edges, indicating two different types of zigzag edges, as expected based on the geometrical consideration above. One type of edge shows spectra with a density of states that has inverse-square-root singularity inherent to a 1D system (see supplementary section II) indicating the presence of a hole-like band with a maximum at +183 meV (red line in Fig. 1c). The broadening used in fitting the spectra in Fig. 1C is remarkably small (6meV) for an electronic state that is far from the Fermi level, as compared to other surface electronic states probed using STM (see supplementary section II), suggesting that quasi-particle excitations in this 1D state are long lived.

We identify this type of edges with the type A edge described above, for which the outermost atom at the edge is weakly coupled to the substrate, and for which model calculations indicate the preservation of a subset of 1D topological edge modes of a freestanding bilayer (see supplementary section I). The details of how this hole-like characteristic is consistent with topological nature of expected edge mode are described in more detail below. The other type of edge, which shows no sign of any singularity (blue line in Fig. 1c), is identified as a type B edge, which we expect to have stronger hybridization of its edge electronic states with the Bi layers underneath (see supplementary section I). Further evidence for this assignment can be found by examining STM conductance maps for a hexagonal pit-like defect (Fig. 1d), which shows the 1D edge state singularity (corresponding to the peak at +183 meV) at alternating edges going around the hexagonal perimeter. Geometric consideration for a hexagonal defect with zigzag edges on this surface (pit or island, as in Fig. 1a) shows that it has alternating A & B type edges, hence showing the un-hybridized edge mode (type A) with a 1D density of states' inverse-square-root singularity on three non-adjacent edges of the hexagonal defect.

Having identified that the type A edges show clear signatures of a 1D edge state, we probe this type of edges in more detail to understand its properties and their connection to the "bulk" electronic states of Bi-bilayer (albeit modified by coupling to the bulk three-dimensional (3D) Bi crystal underneath). STM spectroscopic maps obtained along a line perpendicular to the edge as well as measured exactly along the edge are shown in Fig. 2a and Fig. 2c. As shown in Fig. 2a, spectra far from the edges show features associated with the Bi(111) surface electronic properties, which constitute the "bulk" properties of our Bi-bilayers. Approaching the step edge, these features are shifted in energy due to loss of 2D translational invariance when approaching the edge[22]. In contrast to these features is the discontinuous appearance of the 1D singularity (at +183 meV, see Fig. 2a) exactly at the edge, which constitutes the unique electronic property of the edges. We emphasize that the edge state features in the spectra do not move away from the edge as a function of energy (in a range of energies from 35 meV to 183 meV), thereby demonstrating that they are not due to surface states' quasi-particle interference (QPI) typically observed near step edges on metal surfaces[23].

While confined to a few lattice spacings in the orthogonal direction, spectroscopic linecuts along the step edge (Fig. 2c) show the extended nature of these 1D states in the parallel direction. In addition to a peak at +183 meV corresponding to the singularity, ripples characteristic of a dispersing state are observed at energies below the singularity. Combining all this information, as well spectroscopic line-shapes discussed above (Fig. 1c), it is clear that the type A edges of Bi-bilayer island show features associated with propagating 1D edge states similar to that predicted for free-standing Bi-bilayers[4,18].

To determine whether the 1D edge states of Bi-bilayer islands have the predicted topological properties of freestanding Bi-bilayers, we further examine their scattering properties. The finite extent of the bilayer edges leads to coherent scattering at the ends of the straight sections and gives rise to signatures of quantum interference of the edge-mode quasiparticles in spectroscopic linecuts along the edges (Fig. 2c). The 1D Fourier transform of such linecuts shown in Fig. 3 reveals in fact two scattering interference wavevectors ($q=k_F-k_I$) that disperse as a function of energy. One of these scattering wavevectors, $q_1$, appears starting at +183 meV and disperses downward with energy, and is consistent with the one originating from the 1D edge state with hole-like character, as identified earlier in Fig. 1c (with a singularity at +183meV). The second wavevector, $q_2$ can be identified with the projection of 2D surface state on the direction of a step edge (see below and supplementary section V for details).

The scattering wavevector $q_1$ can be understood within a tight binding model of a freestanding Bi bilayer, provided that we include a Coulomb correction to adjust the position of edge state singularity to match the experimentally measured band edge at +183 meV (Fig. 4a, see supplementary section III for the details of the calculation). As shown in the inset of Fig. 4a, the topological nature of the two edge modes, reflected in their spin properties, strongly suppresses backscattering ($q^*$ scattering channel) and only allows scattering between the states of similar spin ($q_1$ scattering channel)[24]. The dispersion of the allowed $q_1$ wavevector from this model calculation (Fig. 4a) not only matches the experimentally measured dispersion of $q_1$ (Fig. 3) remarkably well but also reproduces the suppression of these scattering processes at lower energies. This suppression of $q_1$ scattering signal results from diminishing overlap between initial and final states with different k values and is related to the k-dependent penetration depth of the edge state[18] as well as its spin (see supplementary section III) and orbital textures. This model calculation indicates the overlap of the edge state with those of the bilayer bulk states at energies below about +30meV, which captures our experimental observation that all the edge spectroscopic signatures are modified below this energy range (see Figs. 1c, 2a & 2b). Overall, the strong suppression of backscattering wavevector ($q^*$ in Fig. 4a, which is absent in data on Fig. 3) together with correspondence between our model calculation for a topological bismuth bilayer and experimentally characterized properties of $q_1$ scattering channel confirm that type A zigzag edges in our experiments behave similar to those predicted for a free standing Bi-bilayer quantum spin Hall system.

Finally, a key property of the topological edge state is that it develops within the energy range of the bulk energy gap. To demonstrate that our results are also consistent with this characteristic of topological states, in Fig. 4b we plot the projection of the *ab initio* calculation of the surface band structure calculation for Bi(111)[20,21] along the momentum direction of our zigzag step edges. This surface band structure, which fits the existing angle-resolved photoemission experiments on Bi(111) surface[17,25] below the Fermi energy ($E_F$), can also be validated above $E_F$ when compared with the STM experiments from surface point-like defects away from the step edges (see supplementary section IV). More specific to our edge properties, this band structure can also be used to understand the origin of the $q_2$ feature of QPI interference shown in Fig. 3 as originating from projection of surface state QPI on the direction along the step edge (see supplementary section V). In effect, this band structure reflects the bulk properties of our Bi-bilayer islands, which are not fully gapped since they are coupled to the underlying bulk Bi crystal. Although not insulating, this surface band structure can be viewed as having a momentum-dependent energy gap, which provides a momentum-energy window within which the 1D topological edge states derived from a freestanding bilayer can coexist with the metallic surface states.. As anticipated from theoretical calculations (see supplementary section I), within that window a subset of the topological edge state survives hybridization with the bulk and remains localized to within a 1D edge channel on type A edges. As shown in Fig. 4b, a combination of the energy of the 1D singularity and the dispersion of $q_1$, which can be understood within our tight-binding model of the zigzag edges (Fig. 3 & 4a), constraints the energy-momentum structure of a subset of 1D edge band to reside within the momentum-dependent gap obtained from the *ab initio* calculation[20,21].

The fact that the type A edge states are weakly coupled to either bulk or surface states of the Bi crystal and are protected from backscattering naturally results in its highly coherent quasi-particle properties. This coherence not only manifest itself in the small intrinsic broadening as reflected in the sharp point spectra along the edge (Figure 1c) but it also results in possibility to resolve very small energy-level quantization of the edge state in restricted geometries. Figure 5 shows differential conductance maps of a type A step edges along a 400Å long edge at different energies together with the line cuts from these data at the edge as a function of energy. The quantization of the edge modes is apparent through visualization of "particle-in-a-box" like state in both maps and the linecuts. The quantization levels of these 1D states are formed between the propagating states of the same "spin" branch, namely through interference in the $q_1$ scattering channel, and is fully consistent with the topological nature of these states (similar to size-quantization effect previously seen on topological surface states of Sb[22]). As the data in Fig. 5 shows, changing the bias by only 3meV (close to our experimental energy resolution) we can clearly resolve the change of the profile of interference associated with size quantization. This observation provides an upper bound for the intrinsic broadening of these states, since the energy level broadening can be influenced by the edge ends, where the edge mode would be hybridized with the states outside of the edge channel. Our ability to resolve these energy levels demonstrates that the propagation of our 1D edge state along the edge must be coherent and is consistent with that expected from topological edge mode that is decoupled from bulk states and protected from backscattering.

While more detailed calculations are required to fully capture the properties of our bilayer islands, the combination of theoretical and experimental results presented here clearly indicates that the edge states of our bilayers have all the predicted topological properties of the edge states of a free-standing Bi-bilayer. We note that the presence of such topological 1D edge states on type A edges of Bi-bilayer does not imply that bulk Bi crystal is topologically non-trivial. Whether or not bulk sample is topological depends on whether it possesses robust surface modes when considering surfaces of bulk sample without a boundary. Such a consideration renders 3D Bi crystal topologically trivial.[26,27] In our experiment Bi(111) surface consists of bilayers with edges that possess edge modes and, as we have argued here, the origin of these modes can be traced back to the edge modes of free standing Bi-bilayer predicted by Murakami.[4] If the coupling of the top bilayer island to the substrate were reduced the surface bands in Figure 4B would develop a gap at all momenta, evolving to a structure shown in the inset of Figure 4A, and the edge mode would be decoupled at all momenta from all surface states.

Clearly, further STM experiments can provide more precise characterization of the properties of edge modes in different situations. For example, measurements in magnetic field or addition of magnetic defects near the edges (by *in situ* deposition) can be used to break time-reversal symmetry[28,29] and to examine the localization of the states spatially[30–32]. Contact with superconducting islands can be used to examine superconducting proximity effect in such 1D states and potentially probe presence of Majorana fermions forming at the ends of such edges[33]. Ultimately, if the edge structure of individual Bi-bilayers prepared on insulating substrates can match those we have created in this work by cleaving, these 1D boundary modes can be exploited for realizing various device proposals with topological edge modes.

## Methods

Single Bismuth crystals were grown using the Bridgman method from 99.999% pure Bi that had been treated to remove oxygen impurities. The samples were cleaved at room temperature in ultra-high vacuum conditions and cooled down to the temperature T=4 K at which STM measurements (with a mechanically sharpened platinum-iridium tip) were carried out. The results reported here have been reproduced on multiple cleaves of different Bi samples and have been confirmed to be stable for different microtips. dI/dV spectra were acquired using lockin amplifier at a frequency of 757 Hz and RMS amplitude of 3mV.

## Acknowledgement

The work at Princeton and the Princeton Nanoscale Microscopy Laboratory was supported by the ARO MURI program W911NF-12-1-0461, DARPA-SPWAR Meso program N6601-11-1-4110, NSF-DMR1104612 , NSF CAREER DMR-095242, ONR- N00014-11-1-0635, and NSF-MRSEC NSF-DMR0819860 programs. S. N-P. acknowledges support of the European Community through the Marie Curie fellowship (IOF 302937).

The authors would like to thank Frank Freimuth for providing the results of *ab initio* calculations and Jungpil Seo and Xi Dai for insightful discussions.

## Author Contributions

I.K.D., S. J., S. N-P, and A.Y designed and carried out the STM measurements and their analysis on samples synthesized by H.J. and R.J.C.. A.A., I.K.D., and B.A.B. performed model calculations and related analysis. All authors contributed to the writing of the manuscript

Correspondence and requests for materials should be addressed to A.Y. (yazdani@princeton.edu)

**Figure captions:**

**Fig.1: Edges of Bi-bilayer Islands on Bi crystal surface. a**, Schematics of Bi-bilayer's atomic structure. Upper panel: top view. Lower panel: side view. Type A and type B edges are marked by red and blue lines respectively. Lattice constant of Bi a = 4.5 Å **b**, Topographic image of the Bi(111) single crystal surface. The height of the Bi-bilayer islands is 4 Å (line profile on the right). Cones of different colors indicate positions at which spectra shown in panel **c** are taken. Inset shows a 30x30 Å$^2$ atomically resolved topographic image of a zigzag edge **c**, Point spectroscopy (I = 7 nA, V = 300 mV) at the two different types of edges A (red line) and B (blue line) and on the surface away from the edges (gray line). While the spectrum on the type A edges shows maximum at $E_2$ = 183 meV which can be fitted to the expected 1D density of states (dashed red line) the spectrum on the type B edge is practically featureless. Point spectrum away from the edges can be fitted using 2D density of states with the maximum corresponding to $E_1$ = 213 meV saddle point in the surface state dispersion (dashed gray line). **d**, Topography close to the hexagonal diatomic depression false colored with differential conductance at $E_2$ = 183 meV (I = 3.5 nA, V = 183 mV). High conductance (red color) is observed at every other edge of a hexagonal pit-like defect.

**Fig.2: Spectroscopic Mapping near the Edge. a**, Spectroscopy across the type A atomic step edge **b,** Topographic linecut across the type A atomic step edge. The spatial resolution of the spectroscopic signatures is limited by the shape of the tunneling apex as well as microscopic details of side-tunneling near the step edge **c**, Spectroscopy along the type A step edge (I = 7 nA, V = 300 mV). **d,** Topographic linecut along the type A step edge. The averaged background conductance at each energy has been subtracted in panels **a** and **c**.

**Fig.3: Quasi-particle Interference within the Edge Channel.** Spatial 1D Fourier transform of the conductance map taken along the type A atomic step edge which coincides with ΓK crystallographic direction. Intensity of the spatial Fourier components (normalized by the number of spatial points) is plotted as false color. Momentum range spans the whole 1D edge Brillouin zone (BZ) from 0 to π/a. Two QPI branches marked with $q_1$ and $q_2$ are identified. The $q_1$ branch corresponds to the 1D edge state (see main text for discussion). The $q_2$ branch corresponds to the projection of 2D surface state QPI (see supplementary section V).

**Fig.4: Model of the Bi-bilayer 1D Edge States. a**, Dispersion calculation of the scattering within the 1D edge mode. The size of the markers indicates the wavefunction overlap between initial ($\psi_i$) and final ($\psi_f$) states (shown in the legend). Inset: Dispersion of the 1D edge state as a function of momentum along the edge (k) calculated for Bi-bilayer. Time-reversal conjugate "spin-up" and "spin-down" branches are schematically indicated by red and blue lines respectively (see supplementary section III for the spin texture). Bulk states are marked with gray. Possible scattering vectors are marked with $q_1$ and q*. The q* vector corresponding to backscattering is prohibited by spin selection rule and is strongly suppressed in the experiment. **b**, Schematics of the quasiparticle dispersion for Bi-bilayer on bulk Bi(111). Points are obtained by projecting the result of *ab initio* calculation of the surface state dispersion[20] onto the 1D BZ of the zigzag edge. Grey region schematically represents the projected Bi(111) surface state continuum on the direction parallel to the edge state. The 1D edge state that shows interference at q* consistent with our data (marked with the red line) exists within the momentum dependent gap defined by the continuum of projected surface states. Momentum transfers corresponding to the red line are obtained from the experiment.

**Fig.5: Size Quantization Effect in Bi Edges. a,** Topographic image of a 400Å long type A zigzag edge of a Bi BL terrace **b-f,** Differential conductance maps at five representative energies resolving how the first five size quantization levels develop within a 1D edge channel of a finite spatial extent. **g,** Linecuts along the edge obtained along the dashed line in panel a showing a typical "particle-in-a-box" behavior. Spectra are offset for clarity.

**Figures:**

**Fig.1**

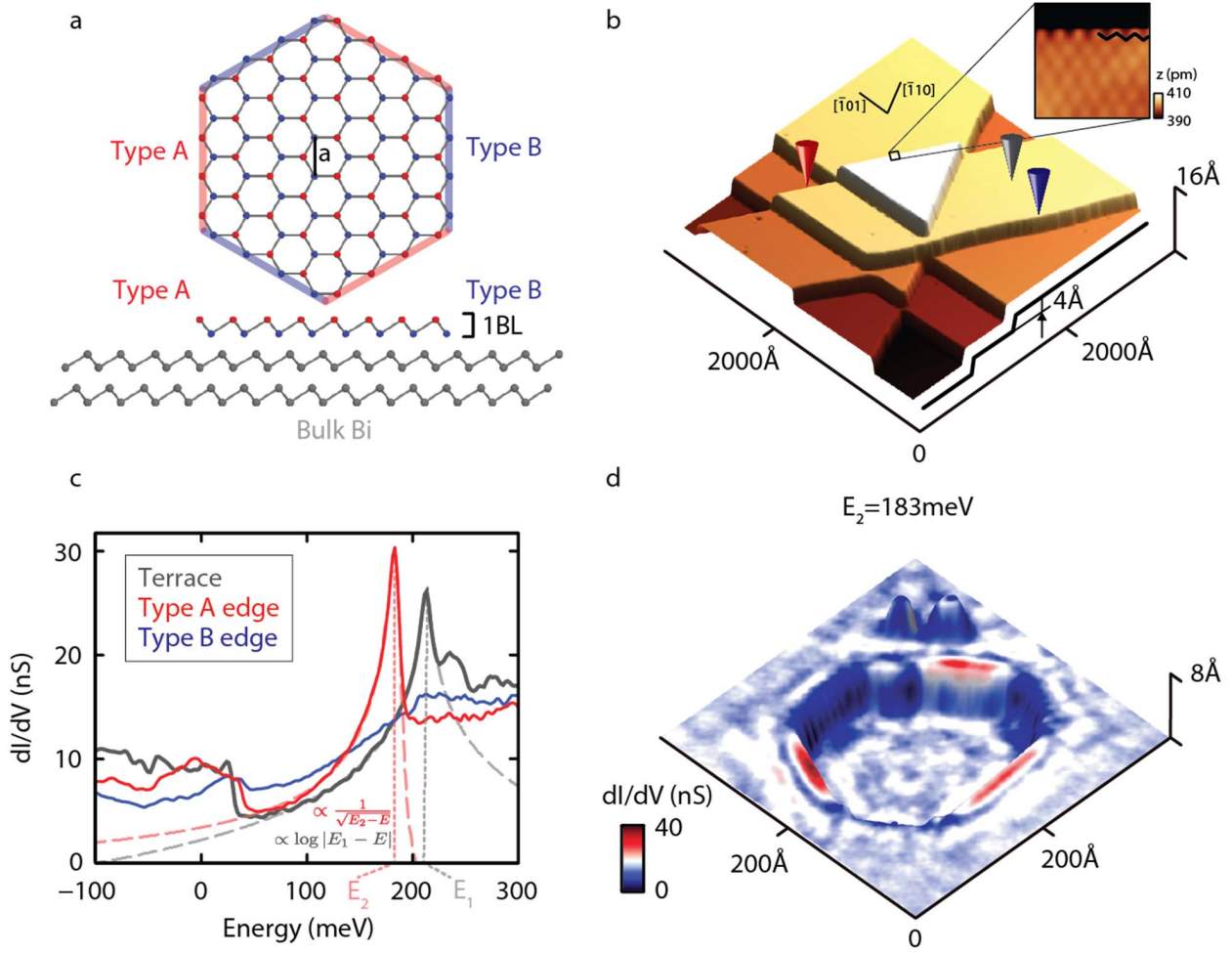

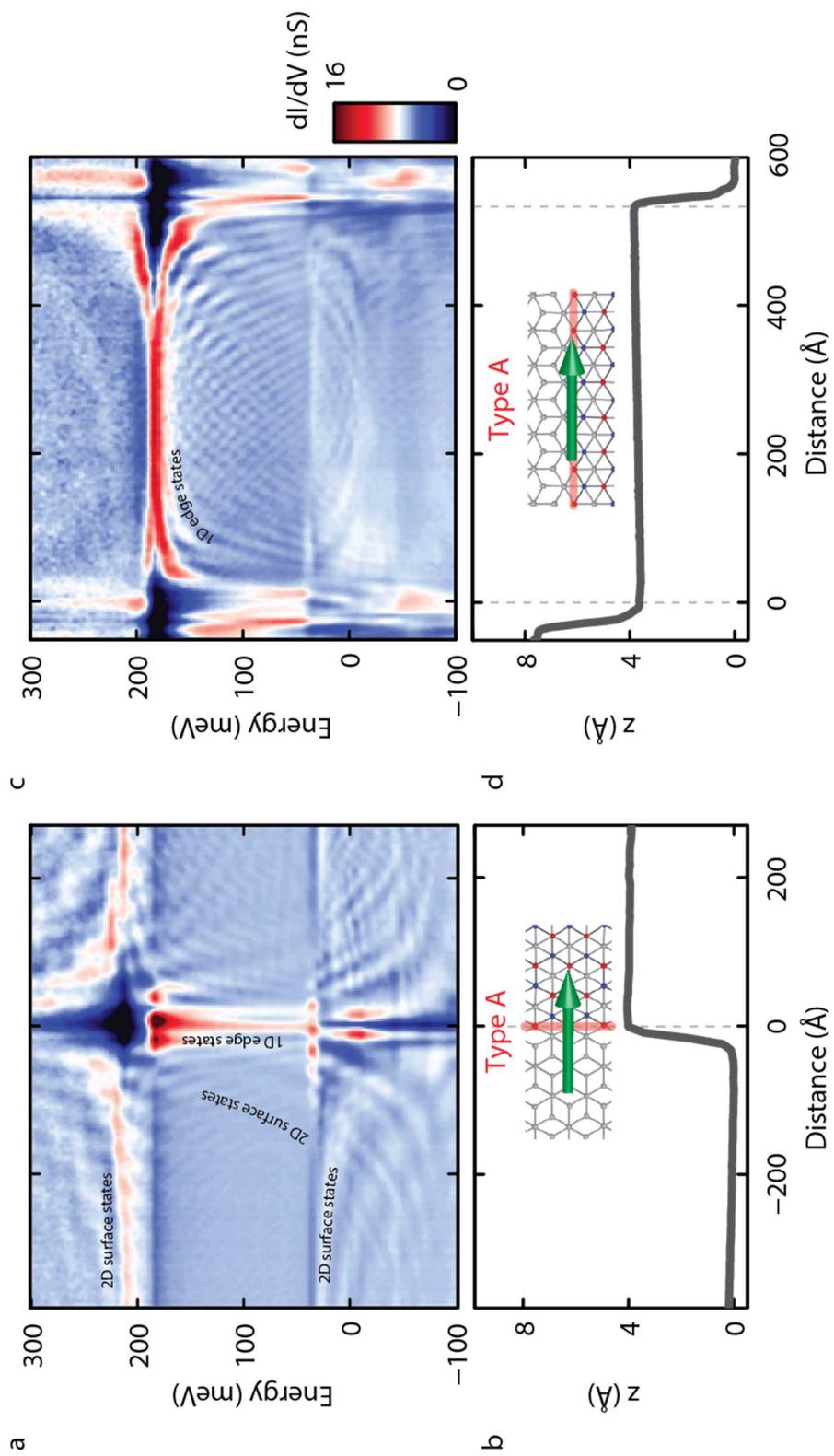

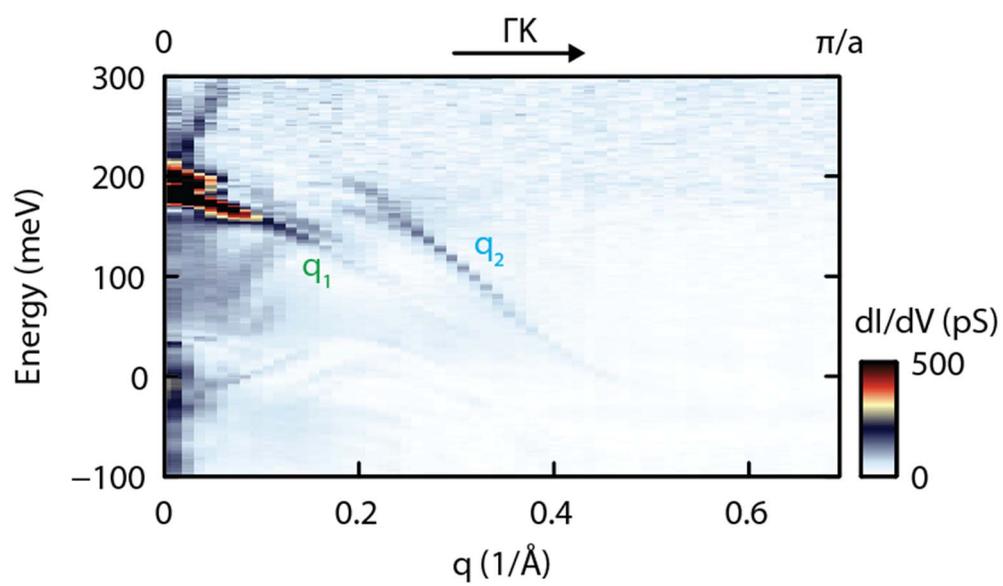

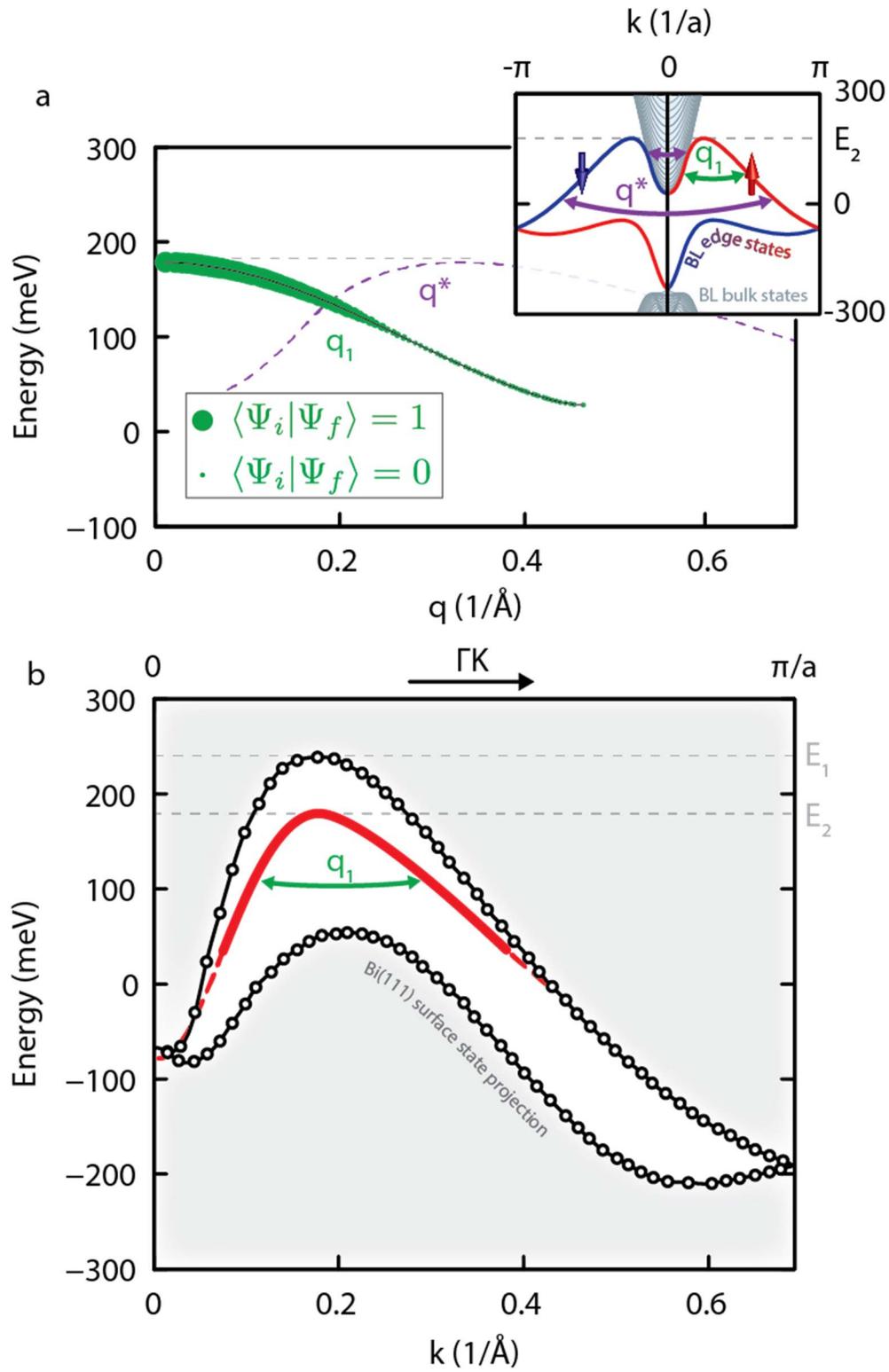

**Fig.5**

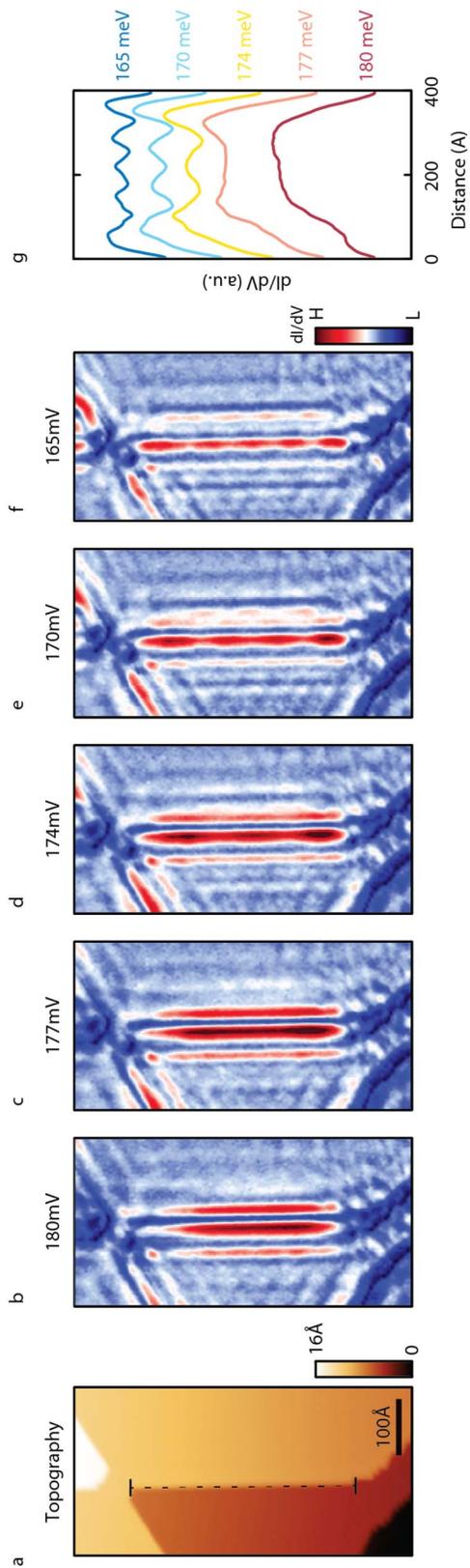

**Supplementary information for "One-dimensional Topological Edge States of Bismuth Bilayers"**

Ilya K. Drozdov[1*], A. Alexandradinata[1*], Sangjun Jeon[1], Stevan Nadj-Perge[1], Huiwen Ji[2], R. J. Cava[2], B. A. Bernevig[1], and Ali Yazdani[1]

[1]*Joseph Henry Laboratories & Department of Physics, Princeton University, Princeton, NJ 08544, USA*

[2]*Department of Chemistry, Princeton University, Princeton, NJ 08544, USA*

**Section I.  Hybridization of Type A and B Edges with the Substrate**

To compare the hybridization of the two different types of zigzag edges with the substrate  we performed a transfer matrix calculation based on the Liu-Allen tight-binding model.[S1] A Bi bilayer ribbon was terminated by two different types of zigzag edges. The 50-unit-cell wide ribbon was placed on top of a 100-unit-cell wide Bi substrate (Fig. S1a).  Periodic boundary conditions were imposed along the Y direction, thus momentum along the edge is conserved. The bulk extends semi-infinitely in the –Z direction. In the transfer-matrix method we match the evanescent eigenfunctions of the Hamiltonian to the boundary conditions imposed by the Bi bilayer ribbon.[S2, S3] These evanescent solutions are non-Bloch states with energies within the gap; these solutions decay exponentially in the –Z direction. The calculation reveals two energetically-distinct modes within the gap, and their wavefunctions are localized to type-A and type-B edges respectively (Fig. S1b). The weight of the edge state on the outermost row of atoms is plotted as a function of the conserved momentum along the edge (Fig. S1c) showing that the type-A edge states are more robust against hybridization with the bulk and surface states than those of type-B edge.

Qualitatively this result can be understood by considering the number of bonds connecting the edges to the bulk. The type-B edge has direct hoppings to the substrate which leads to stronger hybridization of the 1D edge mode with the 3D bulk continuum and results in delocalization. On the other hand, the type-A edge has less bonds with the substrate and thus mimics the edge of a freestanding Bi bilayer.

While the above calculation sufficiently demonstrates the difference in hybridization of the two types of edges, it is known that the Liu-Allen tight-binding model cannot quantitatively reproduce the surface state dispersion.[S4] To quantitatively support our claims, we compare our results together with existing fully-relativistic first-principles calculations in ref. S5, where narrow Bi nanoribbons placed on top of a strained bilayer were investigated. Even though the details of the simulation geometry do not exactly match our experiment (which corresponds to larger Bi islands coupled to a macroscopically large substrate), the calculation of ref. S5 nevertheless predicts the same salient properties of the edge modes as described in the main text, namely: (i) robustness of type-A edge states to hybridization with the substrate, (ii) their localization to within a few unit cells from the edge of the terrace, and (iii) the hole-like singularity $E_2$ in the dispersion.

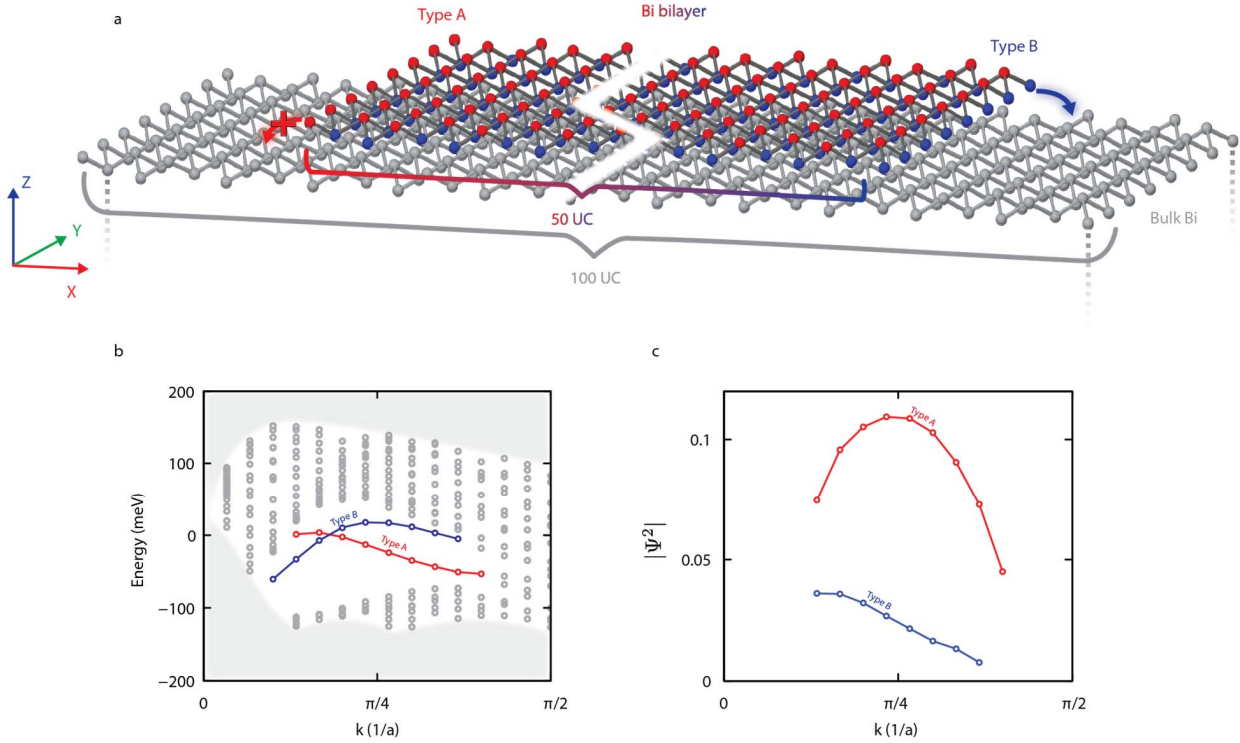

**Fig. S1. Hybridization of type A and B edges to the bulk Bi substrate. a,** The simulation geometry is schematically illustrated. The top atomic layer of the Bi flake are colored red, while the next-to-top layer is colored blue; the underlying Bi substrate is colored grey and extend semi-infinitely in the –Z direction. **b,** Dispersion of surface states, as calculated from the transfer-matrix method with the illustrated geometry. Red (blue) circles correspond to edge states which are localized at the type-A (-B) edge. Grey circles correspond to 2D surface states; the shaded region corresponds to a continuum of 3D bulk states. **c,** Red plot: weights of the type-A edge states, on the outermost atomic row of type A (the left-most, red-colored row in **a**). The weights are plotted as a function of momentum along the edge. Blue plot: weights of type-B edge states, on the outermost atomic row of type B (the right-most, blue-colored row in **a**).

For illustration, we have reproduced some results from Ref. S5 in Fig. S2. In Fig. S2a, we point out that only a subset of the type-A edge mode lies within the energy gap of the Bi (111) surface continuum. We expect that only this subset is robust against hybridization with the substrate of our experiment; the type-B edge mode lies within the surface continuum and is expected to delocalize. To further support this hypothesis, we have also reproduced in Fig. S2b the real-space probability distributions of both types of edge states. Evidently, the type-B edge state extends deeper into the single-bilayer substrate, and is expected to eventually delocalize for thicker substrates. The results of Fig. S2b compare favorably with our theoretical predictions in Fig. S1c.

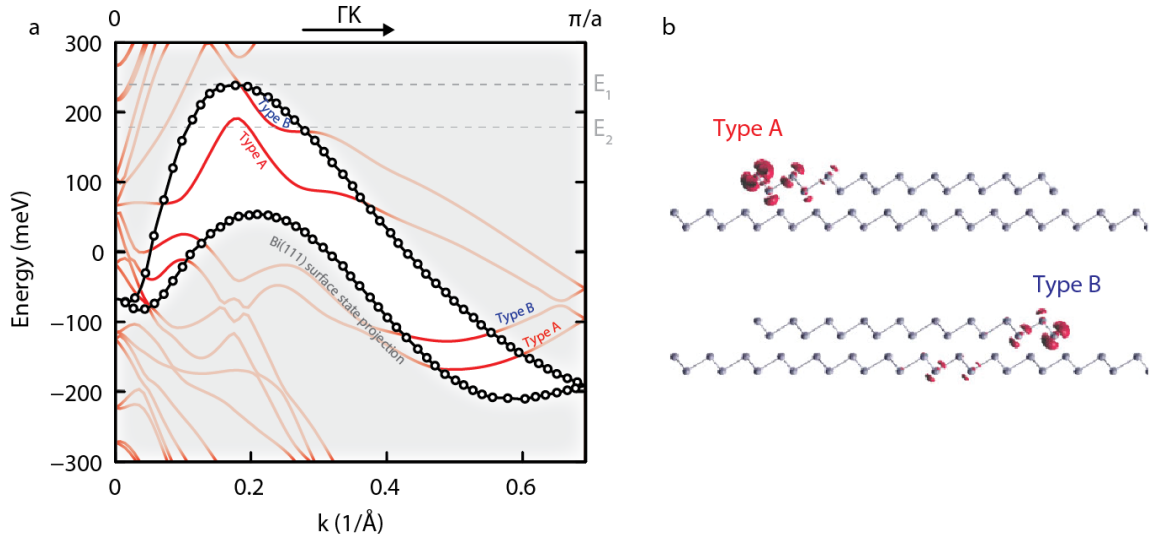

**Fig. S2. Comparison to *ab initio* calculations of Bi nanoribbons placed on top tensile-strained Bi BL. a,** Result from ref. S5 for the two edge modes of a zig-zag edged Bi BL nanoribbon placed on top of Bi BL on Si(111) are plotted in red. Points are obtained by projecting the result of *ab initio* calculation of the surface state dispersion[S6] onto the 1D BZ of the zigzag edge. Grey region schematically represents the projected Bi(111) surface state continuum on the direction parallel to the edge state (same as in Fig. 4 in the main text). Only a subset of edge modes residing within the gap is expected to survive hybridization with the bulk electronic structure. **b,** Wavefunction weight distribution on the two types of edges for the states at momentum $4\pi/10a$ reproduced from ref. S5.

## Section II. Van-Hove Singularities in the Point Spectra & Lifetime of 1D states

In our experiments, we have identified two peaks in the point spectra as arising from two different types of van Hove singularities.[S7] Singularities in the density of states (DOS) typically arise in the vicinity of critical points of the band structure. The functional dependence of the DOS in the vicinity of such critical point depends both on the dispersion of the states, as well as on dimensionality of the system. For example, the inverse-square-root dependence as a function of energy $DOS \propto 1/sqrt(E_0-E)$ is characteristic of 1D parabolic dispersion with band maximum occurring at $E_0$.

The point spectrum away from the step edges on Bi(111) surface in Fig. 1c displays a prominent peak at $E_1$=213meV, which is identified with a Lifshitz transition in the Bi(111) surface state dispersion (Fig S3a). This transition is a change in the topology of the constant energy contours (CECs) from Fig. S3b to Fig. S3c, and the critical point is a saddlepoint in the 2D dispersion. The DOS is expected to diverge logarithmically at a saddlepoint, thus we fitted the DOS near $E_1$ with the formula $a + b \log |E-E_1|$.

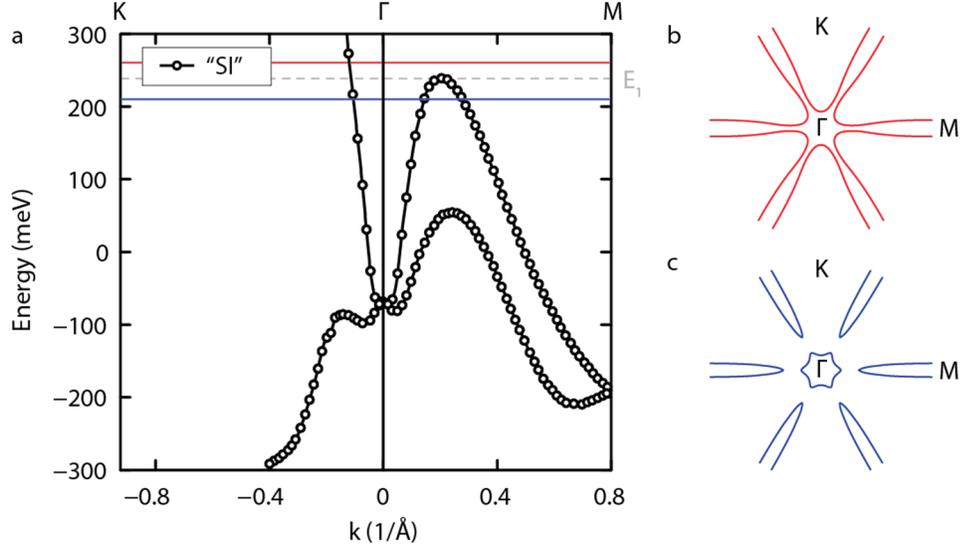

**Fig. S3. Bi(111) surface state dispersion. a,** The results of *ab initio* Bi(111) surface state calculations are plotted along the high-symmetry directions in the 2D BZ (data reproduced from ref. S6). **b, c** Schematic of the two different CEC topologies of the Bi(111) surface state above and below $E_1$. The schematics were drawn based on the results reported in S8 for a 10BL calculation.

In contrast, the point spectra on type-A edges show an asymmetric peak around $E_2$=183meV, which does not correspond to any critical points in the surface states' band structure. The spectrum around $E_2$ is fitted by a + b/sqrt($E_2$-E) function with a small $\Gamma$=6meV imaginary component to energy, which represents all the experimental as well as intrinsic energy broadening of our spectroscopic measurements of these 1D states. The resulting fit is plotted as red dashed line in Fig 1c. A large portion of the observed broadening is experimental, since it should consist of 1.3meV ($3.5k_BT$) of thermal broadening at T=4K and broadening due to finite excitation (3mV RMS), resulting in a total experimental resolution of about 4.5meV. Remarkably, the observed broadening is far smaller than observed before in any type of surface or edge state at energies so far from away from the chemical potential (in this case 200meV). Usually at such energies electron-electron scattering would render electronic states broadened even if they were not coupled to any other states such as bulk states. We can for example contrast this broadening to 50meV broadening of 2D topological surface states of Sb measured by STM[S9]. In contrast, the edge states of Bi exhibit much longer quasiparticle lifetimes, even despite the presence of a conducting substrate.

## Section III. Phenomenological Model of the Topological 1D Edge States

In order to compare experimental data with the predicted QSH edge state of a freestanding Bi bilayer, we employ a Liu-Allen tight-binding model, which we modify with a self-consistent Hartree term to account for Coulomb screening along the edge[S10]. By fitting just a single parameter (the Hubbard on-site energy) we can match the hole-like singularity of the calculated edge mode (inset Fig. 4a) to the experimentally observed hole-like singularity at $E_2$=183meV (Fig. 1c).

This simple model, matching rather well with the experimental data, also captures the essential physics expected from QSH edge states: (i) the odd number of

left- (right-) movers at Fermi level (ii) absence of backscattering between the time-reversed states (iii) momentum-dependent penetration depth of the edge states[S11]. Specifically with regard to Bismuth, the model produces a hole-like parabolic dispersion that is consistent with our experiment. To compare with QPI data, we compute the overlap of wavefunctions at equal energies and different momenta. Away from the parabolic maximum, the overlap diminishes with increasing momentum separation (Fig. 4A, main panel), as is consistent with experimental data (vanishing intensity of $q_1$ mode as a function of increasing momentum transfer in Fig. 3).

In Fig. S4b the spin expectation values for a freestanding bilayer model are plotted for the top branch of the edge dispersion (Fig. S4a). The degree of spin polarization monotonically depends on the momentum along the edge. At the highest momenta, the degree of spin polarization is comparable to that of the Bi(111) surface states.[S12]

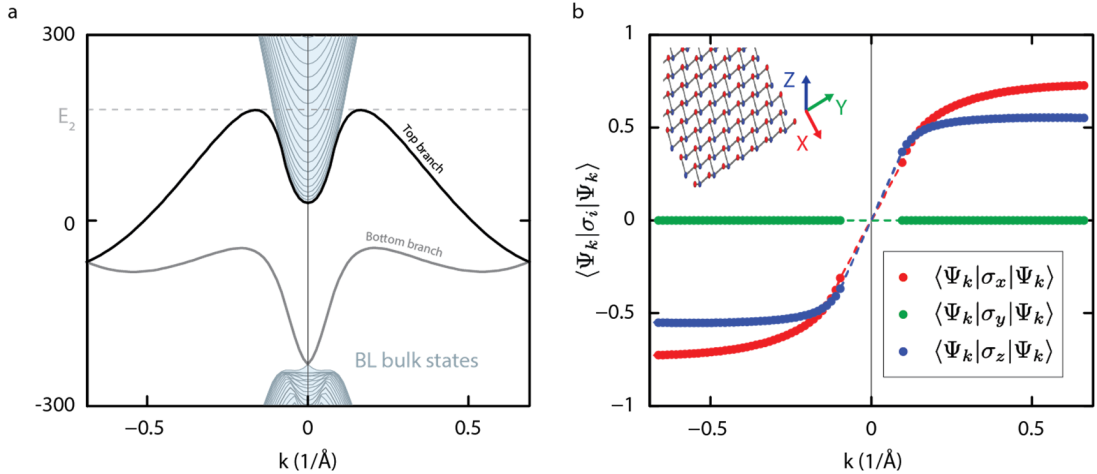

**Fig. S4. Dispersion and spin texture in a freestanding bilayer model. a,** Dispersion of the zigzag edge state of a freestanding Bi BL. **b,** Spin expectation values for the top branch of the zigzag edge state dispersion are plotted as a function of momentum along the edge. Inset shows the orientation of the coordinate basis with respect to the edge.

## Section IV. Probing 2D Quasiparticle Interference on the (111) Surface of Bi

Due to energy overlap between the semi-metallic surface state of Bi (111) and the 1D edge state band it is important to reliably separate the 1D edge signatures from the 2D surface state contribution. For this purpose we have experimentally studied surface state 2D quasiparticle interference away from the step edges. Iron adatoms acting as point-like scatterers were deposited *in situ* on a clean (111) surface of Bi away from the edges. 2D Fourier transforms of real space conductance maps at different energies (Fig. S5a) reveal the elastic scattering processes allowed by spin-selection rules[S13]. Fig. S5b shows the energy-momentum dispersion of the QPI modes nested along the two high-symmetry (ΓM and ΓK) directions in the 2D BZ in the same energy range in which the 1d spectroscopic linecuts (in Fig. 2 and Fig. 3) were acquired.

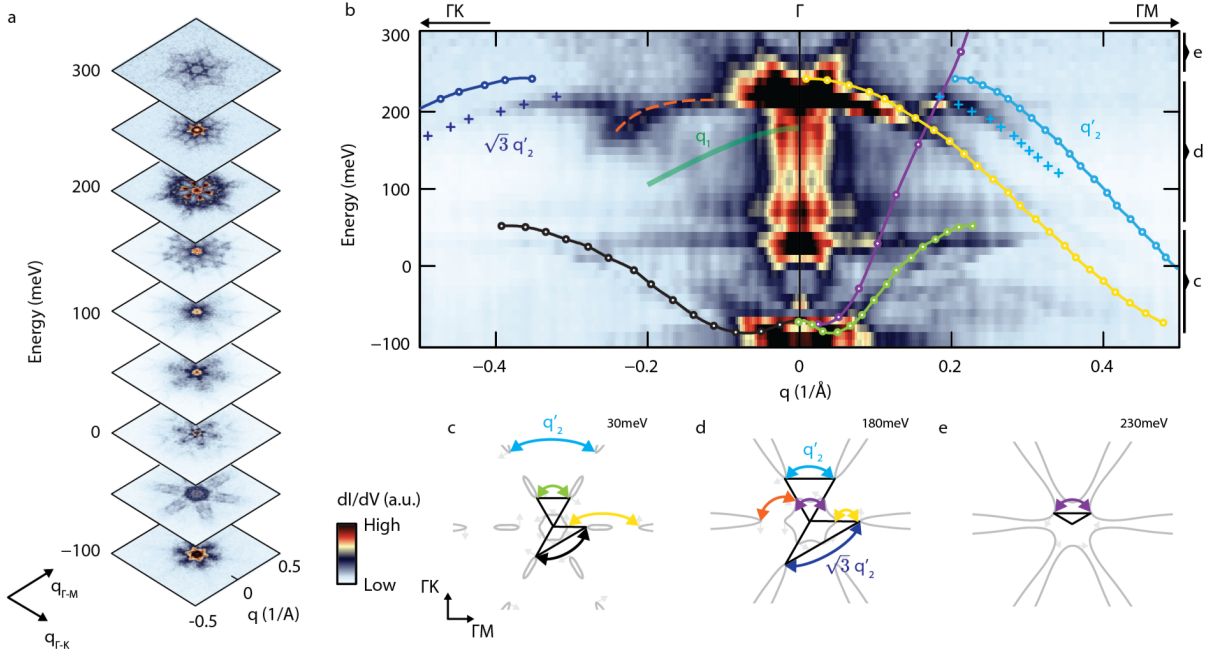

**Fig. S5. Bi (111) surface state quasiparticle interference measurements**. **a,** 2d Fourier transform of real space conductance maps at few representative energies **b,** QPI branches nested along the ΓK and ΓM directions. Momentum differences for the QPI modes obtained from *ab initio* bandstructure[S6] (solid lines and circles) are plotted on top of experimental data. Crosses mark the experimental dispersion of q'$_2$ mode used for comparison with 1D spectroscopic data in the following. Green guide to the eye marks the 1D edge state dispersion obtained from experiment (Fig. 3). **c-e,** Schematic CECs of three representative topologies are plotted in grey. The schematics were drawn based on the results reported in S8 for a 10BL calculation. Light grey arrows schematically represent the spin texture of the surface state. Colored arrows correspond to different scattering processes allowed by spin selection rules. Color coding of scattering processes is consistent across panels b-e.

The resulting experimental QPI dispersion can be compared to the *ab initio* band structure calculations from ref. S6. By considering CECs and associated spin texture (Fig. S5c-e), nested q-wavevectors allowed by spin selection rules can be identified (arrows in Fig. S5c-e). The corresponding momentum differences are calculated from the theoretical surface band dispersion. The nested scattering wavevectors are approximated by the momentum differences between the tips of the pockets and are plotted on top of experimental data for direct comparison (circles and solid lines in Fig. S5b). The resulting dispersion branches capture well the overall shape of the experimental QPI features with some minor quantitative discrepancy most likely present either due to surface band bending effect or due to slight tip-sample workfunction mismatch.

Finally, a guide to the eye corresponding to the experimental 1D edge state dispersion (obtained from Fig. 3) is plotted on top of surface state 2D QPI showing that no features of 2D QPI nested along ΓK can be attributed to the 1D edge state dispersion described in the main text.

## Section V. Projection of the 2D QPI on 1D Spectroscopic Linecuts

To explain the $q_2$ feature observed in Fig. 3 we consider the projection of the surface QPI modes on the direction of the edge. This measurement is performed in a geometrical configuration shown in Fig. S6a in which surrounding step edge geometry acts an interferometer for the 2D surface states[S9]. Likewise, the surface state QPI that is nested not necessarily along the direction of the edge is projected on the direction of the 1D measurement resulting in a projected feature corresponding to $q_2$ (Fig. S6b). Note that this mode does not smoothly connect to the singularity attributed to the 1D edge state. To prove that this mode is indeed originating from the surface state bands we project the experimental dispersion of the surface state intervalley scattering QPI branch (see supplementary section IV, blue crosses in Fig. S5b) on the direction of the measurement (Fig. S5b) which establishes a good agreement between the two independent measurements and confirms the identification of the $q_2$ feature.

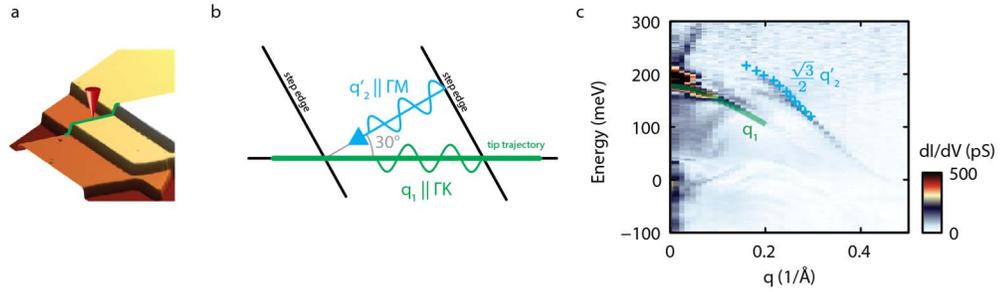

**Fig. S6. Surface state QPI projection. a,** A topographic image of a type-A edge and its surroundings. Green line schematically shows the tip trajectory corresponding to the spectroscopic linecut presented in Fig 2c. **b**, Schematic illustration of how surface state intervalley $q_2$ mode trapped between two parallel step edges gets projected onto the direction of the measurement **c,** Fourier transform from Fig. 3 superimposed with projected experimental dispersion of $q_2$ mode. Blue crosses corresponding to $q'_2$ surface state mode are derived from 2d QPI experiment (Fig. S5b) and are projected on the direction of the step edge with a geometric $\sqrt{3}/2$ factor. Green guide to the eye marks the 1D edge state dispersion.

## Supplementary references